\documentclass[aps,prb,floatfix,twocolumn,nofootinbib,showpacs,superscriptaddress,longbibliography]{revtex4-2}
\usepackage{graphicx}
\usepackage{subcaption}
\usepackage{amsmath,amssymb}
\usepackage{flushend}
\usepackage[justification=raggedright]{caption}
\usepackage[utf8]{inputenc}
\usepackage[colorlinks = true,
linkcolor = blue,
urlcolor  = blue,
citecolor = blue,
anchorcolor = blue]{hyperref}
\usepackage{array}
\usepackage{bbold}
\usepackage{float}
\usepackage{enumitem}
\usepackage{braket}
\usepackage{appendix}
\usepackage[mathscr]{euscript}
\usepackage{color}
\usepackage{times}
\usepackage{comment}
\usepackage{natbib}

\begin{document}
\title{Competing topological phases in a non-Hermitian time-reversal symmetry-broken Bernevig-Hughes-Zhang model}
\author{Dipendu Halder$^@$}
\email[]{h.dipendu@iitg.ac.in}

\author{Srijata Lahiri$^@$}
\email[]{srijata.lahiri@iitg.ac.in}

\author{Saurabh Basu}
\email[]{saurabh@iitg.ac.in}
\affiliation{Department of Physics, Indian Institute of Technology Guwahati-Guwahati, 781039 Assam, India}

\begin{abstract}
\noindent The Bernevig-Hughes-Zhang (BHZ) model, which serves as a cornerstone in the study of the quantum spin Hall insulators, showcases robust spin-filtered helical edge states in a nanoribbon geometry. 
In the presence of an in-plane magnetic field, these (first-order) helical states gap out to be replaced by second-order corner states under suitable open boundary conditions.
Here, we show that the inclusion of a spin-dependent non-Hermitian balanced gain/loss potential induces a competition between these first and second-order topological phases.
Surprisingly, the previously dormant first-order helical edge states resurface as the non-Hermitian effect intensifies, effectively neutralizing the role played by the magnetic field.
By employing the projected spin spectra and the spin Chern number, we conclusively explain the resurgence of the first-order topological properties in the time-reversal symmetry-broken BHZ model in presence of non-Hermiticity.
Finally, the biorthogonal spin-resolved Berry phase, exhibiting a non-trivial winding, definitively establishes the topological nature of these revived edge states, emphasizing the dominance of non-Hermiticity over the magnetic field.
\end{abstract}

\maketitle
\def\thefootnote{@}\footnotetext{These authors contributed equally to this work}

\section{Introduction}

Over the past few decades, the discovery of the integer quantum Hall effect in the presence of external magnetic field \cite{PhysRevLett.45.494, PhysRevLett.49.405} has sparked a significant increase in the interest in topological phases within condensed matter physics \cite{RevModPhys.82.3045, RevModPhys.83.1057, RevModPhys.88.035005}.
During this period, the fabrication of graphene \cite{Novoselov} spurred significant research into the topological aspects of quantum spin Hall systems \cite{Murakami, PhysRevLett.93.156804} and led to Kane and Mele's identification of the $\mathbb{Z}_2$ topological insulators \cite{PhysRevLett.95.146802, PhysRevLett.95.226801}.
Concurrently, Bernevig, Hughes, and Zhang (BHZ) proposed achieving quantum spin Hall (QSH) effect in HgTe-CdTe quantum wells \cite{BHZ_2006} owing to an electronic band inversion under certain conditions, with its experimental verification following shortly \cite{BHZ_2007}.
This approach eliminates the need for a honeycomb-lattice system and a time-reversal symmetry (TRS) breaking second neighbor hopping found in the original Haldane model \cite{PhysRevLett.61.2015}, thereby enabling 2D topological insulator (TI) states in a broader range of materials with strong spin-orbit coupling, laying the groundwork for further advancement in 2D topological materials.

Recently, the interplay of topology and non-Hermiticity \cite{PhysRevLett.116.133903, PhysRevLett.120.146402, Ghatak_2019, PhysRevX.8.031079, PhysRevX.9.041015, Ashida2020, RevModPhys.93.015005} has emerged as a captivating frontier in the realm of condensed matter physics.
This fortuitous interplay has revealed a plethora of intriguing physical phenomena, prominently featuring the non-Hermitian skin effect \cite{PhysRevLett.121.086803, PhysRevLett.121.026808, PhysRevB.99.201103, PhysRevLett.124.056802, PhysRevLett.124.086801}, where bulk eigenstates primarily localize near system boundaries, exceptional points \cite{PhysRevB.97.121401, Heiss_2012, PhysRevLett.118.040401}, at which the Hamiltonian becomes defective and multiple eigenvectors coalesce, and the non-Bloch band theory \cite{PhysRevLett.123.066404}.
The experimental exploration of the topological systems has been materialized in diverse physical settings, spanning ultra-cold atoms in optical lattices \cite{El-Ganainy2018, Eichelkraut2013}, electronic \cite{PhysRevLett.114.173902, Helbig2020, Lee2018, PhysRevB.109.115407}, mechanical \cite{Wang}, and acoustic \cite{Fleury2015, PhysRevApplied.16.057001} systems.
Thus, non-Hermitian (NH) systems present a vast platform to explore the connection between topology and non-Hermiticity.

The scope of topological insulators has been broadened further to include higher-order topological insulators (HOTIs) \cite{PhysRevLett.110.046404, PhysRevLett.119.246401, PhysRevLett.119.246402, hoti_sc, Schindler2018, Xie2021, PhysRevB.99.041301, PhysRevB.96.245115, Benalcazar}, which host robust topological states at boundaries of dimension $d-2$ or less for a bulk that is $d$-dimensional.
For example, a second-order HOTI hosts zero-dimensional corner modes for a bulk that is two-dimensional and one-dimensional hinge modes for a three-dimensional bulk.
Recent research exploring the interplay between non-Hermiticity and higher-order topological phases (HOTPs) \cite{PhysRevLett.122.076801, PhysRevLett.123.016805, PhysRevB.99.081302, PhysRevLett.123.073601, PhysRevB.102.205118, Qihuang, PhysRevB.106.L140303, ghosh24} suggests the presence of intriguing phenomena yet to be uncovered.
One of the most intriguing puzzles in the field of HOTI involves devising a mechanism to gap out the first-order edge (surface) states of a 2D (3D) TI to transform them into an HOTP.
In this context, Ren et al. \cite{PhysRevLett.124.166804} have demonstrated how an in-plane magnetic field (IPMF) gaps out the helical edge states of a QSH phase by breaking the TRS, thereby giving rise to a HOTP characterized by the presence of distinct corner modes observed in suitable open boundary conditions.
A pertinent question arises: is it feasible to recover the QSH phase from the HOTP? If so, how do we delineate the reentrant QSH phase?

In this work, we employ the BHZ model as an example of a QSH insulator and introduce an IPMF while incorporating non-Hermiticity.
Interestingly, the QSH phase experiences a revival even when the TRS remains broken by the IPMF.
We begin by analyzing the Hermitian BHZ model and show that the inclusion of the TRS-breaking IPMF causes the selective destruction of the helical edge states, regarded as the first-order topological phase (FOTP), in an $x$-periodic (finite along $y$) system, giving rise to a second-order topological phase (SOTP).
However, the helical states remain robust in the $y$-periodic (finite along $y$) system when the strength of the IPMF is small.
We then introduce an NH spin-dependent balanced gain/loss potential in this TRS-broken perturbed BHZ model.
Interestingly, we observe a revival of the helical edge states in the $x$-periodic system as the non-Hermiticity gains dominance over the IPMF.
We establish the physical origin of this competition between the SOTP and the FOTP, with the former characterized by the corner and the latter characterized by the helical edge states, as a function of the NH potential, and decode the establishment of the FOTP in the presence of IPMF in the system.
We also study the spectrum of the projected spin operator and the spin Chern number (SCN) to arrive at a logical conclusion and provide further support for this reentrant behavior.

\section{Model and results}

The BHZ model is a tight binding toy model for a two-dimensional topological insulator, which in its original form exhibits the presence of helical edge modes protected by the TRS and characterized by a $\mathbb{Z}_2$ invariant.
In our work, we study a variant of the BHZ model under different boundary conditions, which can be stated as follows.
(a) \textit{Nanotube configuration}: When the lattice is periodic in both the $x$- and $y$-directions, it is referred to as the `nanotube'.
(b) \textit{Nanoribbon configuration}: When the lattice is periodic in one direction while being aperiodic in the other, it is referred to as the `nanoribbon', with the specific periodic direction mentioned.
(c) \textit{Open boundary condition (OBC)}: When the system is fully aperiodic, it has boundaries in both the directions.
We then construct a rhombic structure which is composed of multiple square BHZ unit cells. We refer to this as the `rhombic supercell' in the text.
Having elaborated upon the different geometric configurations that have been used to study the topology of varying orders in the system, we now embark upon the Hamiltonian for the original (Hermitian) BHZ model \cite{BHZ_2006}, which in the momentum space is given by:
\begin{align}
H(k_x,k_y) = &\left[\Delta-2t(\cos k_x + \cos k_y)\right]\sigma_0\otimes\tau_z + \nonumber \\ &2t_{sp}\left[\sin k_x(\sigma_z\otimes\tau_x)+\sin k_y(\sigma_0\otimes\tau_y)\right]
\label{eq:BHZ}
\end{align}
Here $\Delta$, $t$, and $t_{sp}$ are real-valued parameters with $\sigma$ and $\tau$ being Pauli matrices corresponding to the spin and orbital degrees of freedom, respectively.
Now we include an IPMF, $\pmb{B}=(B_x,B_y,0)$ such that the additional part of the Hamiltonian acquires the form,
\begin{equation}
H_\text{B} = (B_x\sigma_x+B_y\sigma_y)\otimes\tau_0
\end{equation}
The total Hamiltonian $H_T$ is now given by,
\begin{equation}
    H_\text{T} = H + H_\text{B}
    \label{E3}
\end{equation}
It is worth mentioning that, without loss of generality, the magnetic field vector $\pmb{B}$ is aligned along the $y$-direction ($B_x=0$) unless specified otherwise.

Due to the presence of the mirror symmetry $M_y$ in the system, the spectrum of this Hamiltonian is studied on a rhombic supercell, consisting of two $M_y$ invariant corners \cite{PhysRevLett.124.166804}.
The spectrum under such OBC shows zero energy states separated from the bulk. 
It is observed that the probability densities of these in-gap states are confined at the aforementioned corners of the supercell (see Appendix \ref{app1}).
It is also observed that in the presence of IPMF, the helical edge modes of the $x$-periodic nanoribbon gap out.
This indicates that the disappearance of the QSH states induced by the broken TRS results in the emergence of in-gap, zero-energy corner states as a replacement.
However, the edge modes for the $y$-periodic nanoribbon behave differently (see Appendix \ref{app2}).

We now include a spin-dependent imaginary potential of the form,
\begin{equation}
H_\text{NH} = \mathrm{i}\gamma\sigma_z\otimes\tau_0,
\label{E4}
\end{equation}
\begin{figure}[t]
\centering
\begin{subfigure}{\columnwidth}
         \centering
         \includegraphics[width=\columnwidth]{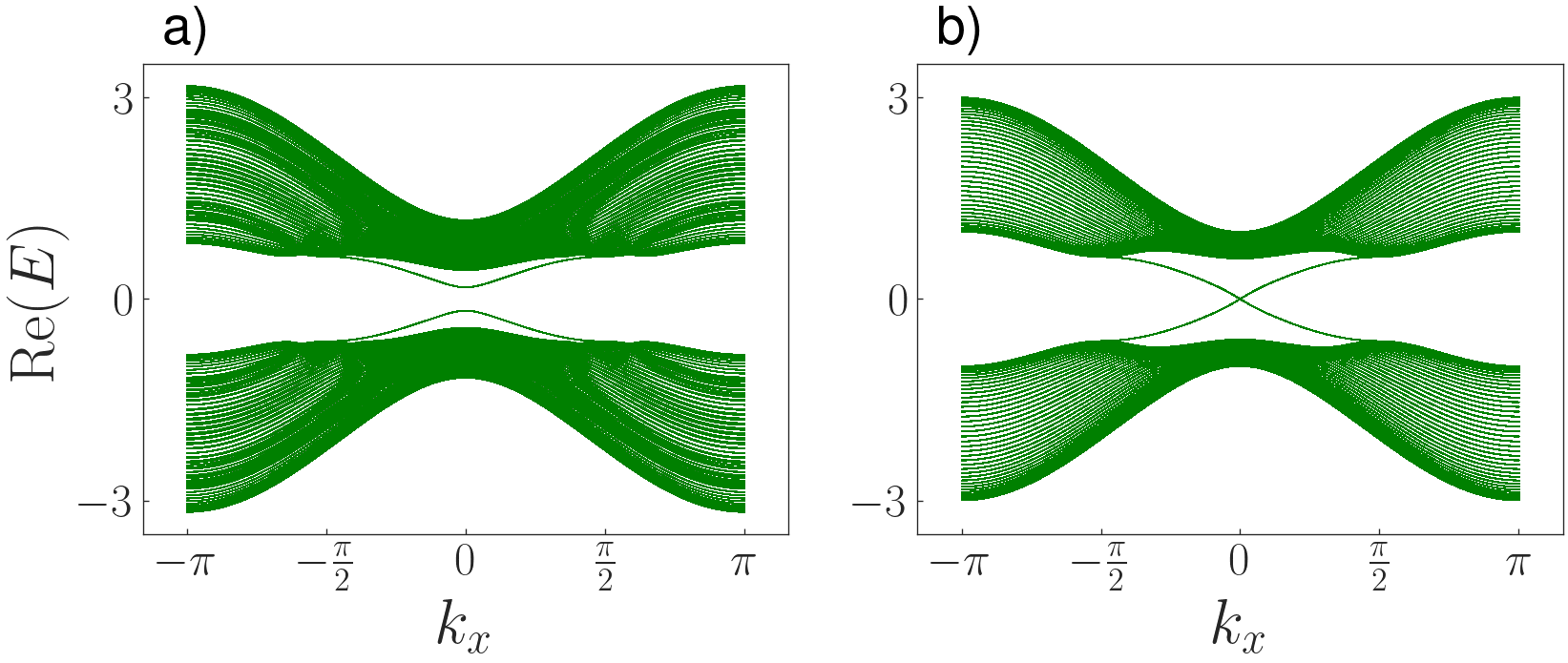}
\end{subfigure}
\begin{subfigure}{\columnwidth}
         \centering
         \includegraphics[width=\columnwidth]{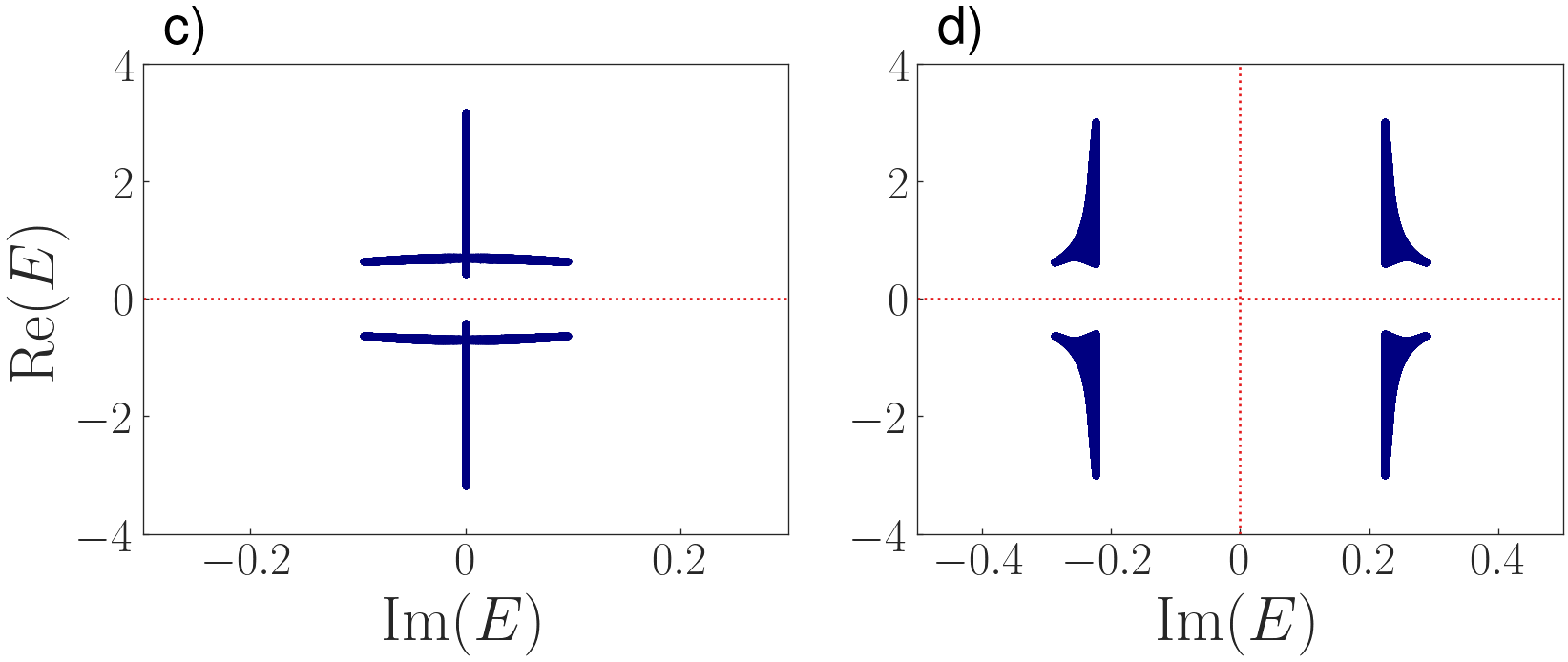}
\end{subfigure}
\caption{The real part of the energy spectra of the $x$-periodic nanoribbon for the Hamiltonian $H_\text{R}$ is plotted. For $\gamma<|B_y|$ ($|B_y|=0.2$), the real energy spectrum is gapped (Fig. \ref{Fig1}(a)). However, the helical states are restored when $\gamma$ exceeds $B_y$ (Fig. \ref{Fig1}(b)). The bulk Re($E$) vs. Im($E$) spectra show a transition from a real line gap (Fig. \ref{Fig1}(c)) to a mixed line gap (Fig. \ref{Fig1}(d)) as a function of the non-Hermiticity. Here the strength of the non-Hermiticity, that is $\gamma$, takes a value of $0.1$ for Fig. \ref{Fig1}(a), (c) and $0.3$ for Fig. \ref{Fig1}(b), (d).}
\label{Fig1}
\end{figure}
where $\gamma$, being a positive real quantity, describes the strength of the non-Hermiticity.
This balanced gain/loss induces an energy disparity between the spin-up ($\uparrow$) and the spin-down ($\downarrow$) states.
When the non-Hermiticity $H_\text{NH}$, described by Eq. \eqref{E4}, is introduced in the BHZ Hamiltonian (Eq. \eqref{eq:BHZ}), the helical edge states along both the $x$ and $y$ directions persist, since $H_\text{NH}$ {\it does not} break the TRS of the system.
However, the energy spectra exhibit pairs such as $E_R\pm i\gamma$, where $E_R$ is the real part of the eigenspectra (see Appendix \ref{app1}).
Notably, a significant amount of experimental work has been conducted concerning spin-dependent potentials within two-dimensional ultracold atomic gases \cite{Lin2011, Galitski2013, PhysRevLett.109.095301, PhysRevLett.112.086401}.

Our primary objective is to investigate the impact of non-Hermiticity on the $x$-periodic nanoribbon in the presence of the IPMF to observe how the edge states, which were gapped out by the IPMF, respond to the non-Hermiticity.
Henceforth, unless otherwise specified, we will refer to an $x$-periodic nanoribbon as a nanoribbon to avoid verbose expressions.
We find that as the strength of the non-Hermiticity $\gamma$ approaches the strength of the IPMF, $|B_y|$, the band gap in the real part of the energy spectra slowly diminishes, as presented in Fig. \ref{Fig1}(a).
At the critical point, $\gamma=|B_y|$, this real energy band gap closes, and thus, the dispersive helical states connecting the conduction and valence bands are revived in the region $\gamma\ge |B_y|$ (Fig. \ref{Fig1}(b)).
Let us denote the resulting Hamiltonian by $H_\text{R}$, which under a nanotube configuration, is obtained via adding $H_\text{NH}$ to $H_\text{T}$,
\begin{equation}
    H_\text{R}=H_\text{T}+H_\text{NH}
    \label{eq:R}
\end{equation}
The spectrum of $H_\text{R}$ illustrates a noteworthy transition: from a real line gap \cite{PhysRevX.9.041015} for $\gamma\leq|B_y|$ (Fig. \ref{Fig1}(c)), to a mixed line gap in the region where $\gamma$ exceeds $|B_y|$ (Fig. \ref{Fig1}(d)).
The spectral gap transition, occurring at $\gamma=|B_y|$, signals the presence of exceptional points in the system, marking the onset of a criticality where the properties undergo significant transformations.

\begin{figure}
\includegraphics[width=\columnwidth]{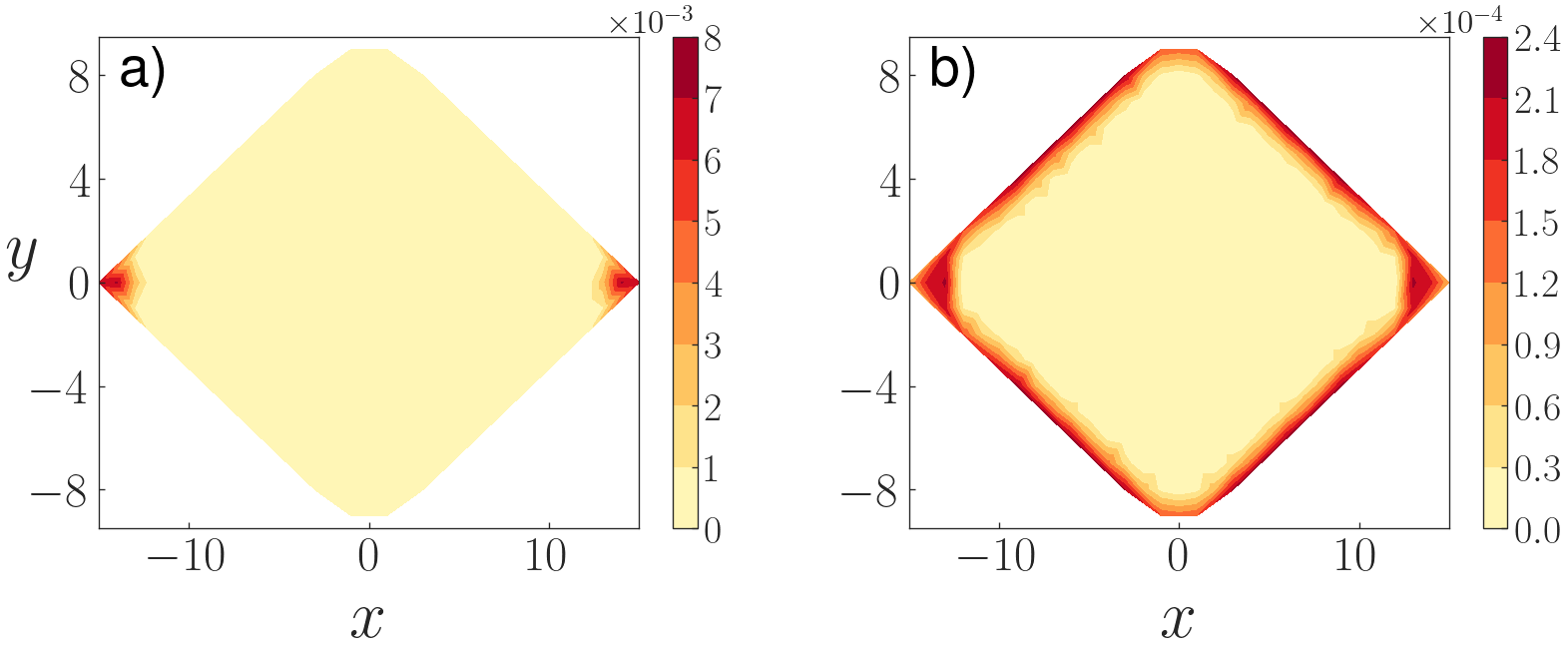}
\caption{(a) The probability distribution of the zero energy states in the NH HOTI phase, that is, $\gamma<|B_y|$, is shown. (b) The probability distribution of the same state in the regime $\gamma>|B_y|$ shows a spread along the edges of the supercell. The value of $|B_y|$ is kept fixed at $0.2$.}
\label{Fig2}
\end{figure}
We now focus on analyzing the Hamiltonian under full OBC.
In the region where $\gamma<|B_y|$ diagonalization of  $H_\text{R}$ on the previously mentioned rhombic supercell shows that the zero energy states similar to the Hermitian case (Eq. \eqref{E3}) still persist.
The bi-orthogonal probability densities of these zero energy states ($\mathrm{Re}(E)=\mathrm{Im}(E)=0$) show confinement at the $M_y$ invariant corners of the rhombus, which confirms the existence of the NH HOTI phase as long as the effect of the IPMF dominates (Fig. \ref{Fig2}(a)). 
Beyond the critical point ($\gamma = |B_y|$), the probability distribution of the states that were earlier confined at the corners of the rhombus now get delocalized along the edges of the supercell (Fig. \ref{Fig2}(b)). 
This situation highlights the re-emergence of FOTP, characterized by the presence of helical edge states.
Hence, we affirm that a very clear tussle exists between an SOTP and an FOTP, which is directly influenced by the strength of the non-Hermiticity.

\section{Analysis of the revival of FOTP}

We now focus on analyzing the physical mechanism behind the suppression of the SOTP and resurgence of the helical edge states in the NH BHZ model in the presence of the IPMF.
Due to the absence of TRS in the system, the conventional strategy of calculating the $\mathbb{Z}_2$ invariant ceases to be useful. 
Hence, we switch to calculating SCN, $C_{\uparrow/\downarrow}$, a valid topological invariant for QSH systems having no TRS \cite{PhysRevB.80.125327, PhysRevLett.107.066602}. 
Our initial emphasis would be on analyzing how the indicator of an FOTP (in our case, the SCN) for the hermitian BHZ Hamiltonian gets affected by the IPMF.
To evaluate $C_{\uparrow/\downarrow}$, we construct a projector, $P(\mathbf{k})$, on the occupied subspace of the Hamiltonian, $H_\text{T}$, presented by the Eq. \eqref{E3},
\begin{equation}
P(\mathbf{{k}})=|u_1(\mathbf{k})\rangle\langle u_1(\mathbf{k})|+|u_2(\mathbf{k})\rangle\langle u_2(\mathbf{k})|,
\end{equation}
where $|u_1(\mathbf{k})\rangle$ and $|u_2(\mathbf{k})\rangle$ correspond to the valence eigenstates. 
In this case, the projector is a $4\times 4$ matrix.
The projected spin operator $S$ is constructed via,
\begin{equation}
S(\mathbf{k})=P(\mathbf{{k}})\hat OP(\mathbf{{k}}),
\label{E6}
\end{equation}
\begin{figure}[t]
\centering
\includegraphics[width=\columnwidth]{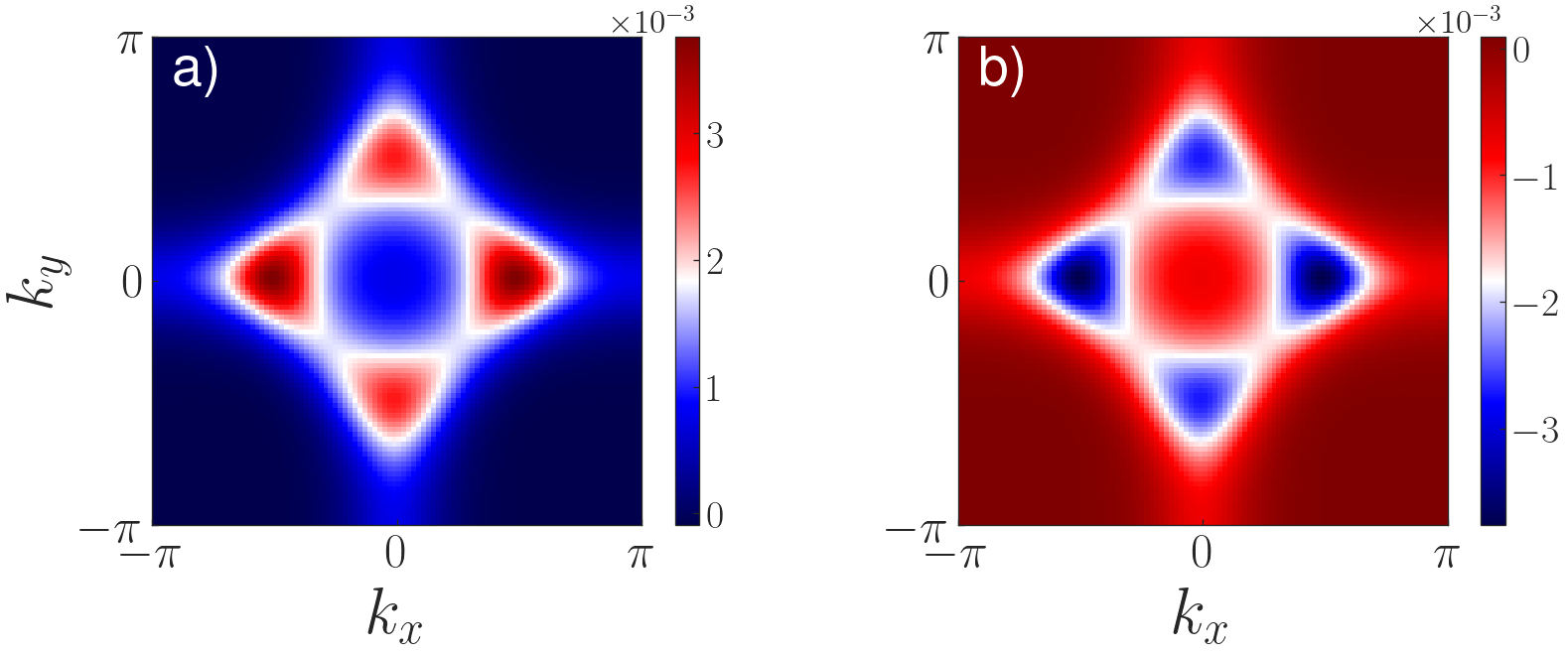}
\caption{The spin Berry curvature for the (a) $\uparrow$-spin and the (b) $\downarrow$-spin states is calculated using the eigenvectors $\zeta_1$ and $\zeta_4$ of the Hermitian-projected spin operator $S(\mathbf{k})$ (Eq. \eqref{E6}), respectively. $t$, $\Delta$, $t_{sp}$ and $B_y$ are kept fixed at $0.5$, $1$, $0.3$ and $0.2$ respectively.}
\label{Fig3}
\end{figure}
where $\hat O=\sigma_z\otimes\tau_0$.
Diagonalization of $S$ gives four distinct eigenvalues, $(E^s_1,E^s_2,E^s_3,E^s_4)$, of which $|E^s_1|=|E^s_4|$ and $E^s_2=E^s_3=0$. 
Employing the Fukui's method \cite{doi:10.1143/JPSJ.74.1674}, the SCNs $C_{\uparrow}$ and $C_{\downarrow}$ are now respectively calculated using the eigenvectors $|\zeta_1\rangle$ and $|\zeta_4\rangle$ of $S(\mathbf{k})$, corresponding to the eigen-energies $E^s_1$ and $E^s_4$.
Further the spin Berry curvature, $\Omega^s_{\uparrow/\downarrow}(k_x, k_y) $ is plotted in Fig. \ref{Fig3} where \cite{doi:10.1143/JPSJ.74.1674, PhysRevLett.61.1329},
\begin{align}
    \begin{split}
       \Omega^s_{\uparrow/\downarrow}(\vec k) &=i\biggl{[} \biggl <\frac{\partial \zeta(\vec k)}{\partial k_x}\biggl|\frac{\partial \zeta(\vec k)}{\partial k_y}\biggl> - \biggl <\frac{\partial \zeta(\vec k)}{\partial k_y}\biggl|\frac{\partial \zeta(\vec k)}{\partial k_x}\biggl>\biggl ]
    \end{split}
\end{align}
Here $\zeta\in \{\zeta_1, \zeta_4\}$ corresponding to the $\uparrow$-spin and $\downarrow$-spin states respectively.
Alternative to Fukui's method, integration of the spin Berry curvature over the entire Brillouin zone (BZ) also gives the SCN ($C_{\uparrow/\downarrow}$). 
We observe that the SCN for the Hamiltonian $H_\text{T}$ is non-trivial, namely, $C_\uparrow=-C_\downarrow=1$, even in the presence of the IPMF. 
This indicates that the IPMF {\it does not} destroy the first-order bulk topology. 
It only provides a channel through which the $\uparrow$-spin state can scatter into a $\downarrow$-spin state.
This scattering leads to the gapping out of helical edge states, consequently giving rise to the SOTP, as observed in the form of corner states on the rhombic supercell.
Nevertheless, the introduction of $H_\text{NH}$ gradually mitigates this scattering effect.
We show this by plotting the spectra of the projected spin operator $S_\text{NH}$ constructed for $H_\text{R}$ over the nanoribbon configuration for different strengths of the non-hermiticity $\gamma$ in Fig. \ref{Fig4}.
The projected spin operator, $S(k)$ presented in Eq. \eqref{E6}, is redefined for the NH systems as,
\begin{equation}
S_\text{NH}(k_x) = P_L({k_x})\hat O P_R({k_x}),
\label{E7}
\end{equation}
where 
\begin{subequations}
\begin{alignat}{2}
P_L(k_x)&=\sum_{n\in N_\text{occ}} |u_n^L(k_x)\rangle\langle u_n^L(k_x)|\\
P_R(k_x)&=\sum_{n\in N_\text{occ}} |u_n^R(k_x)\rangle\langle u_n^R(k_x)|.
\end{alignat}
\end{subequations}
\begin{figure}[t]
\centering
\includegraphics[width=\columnwidth]{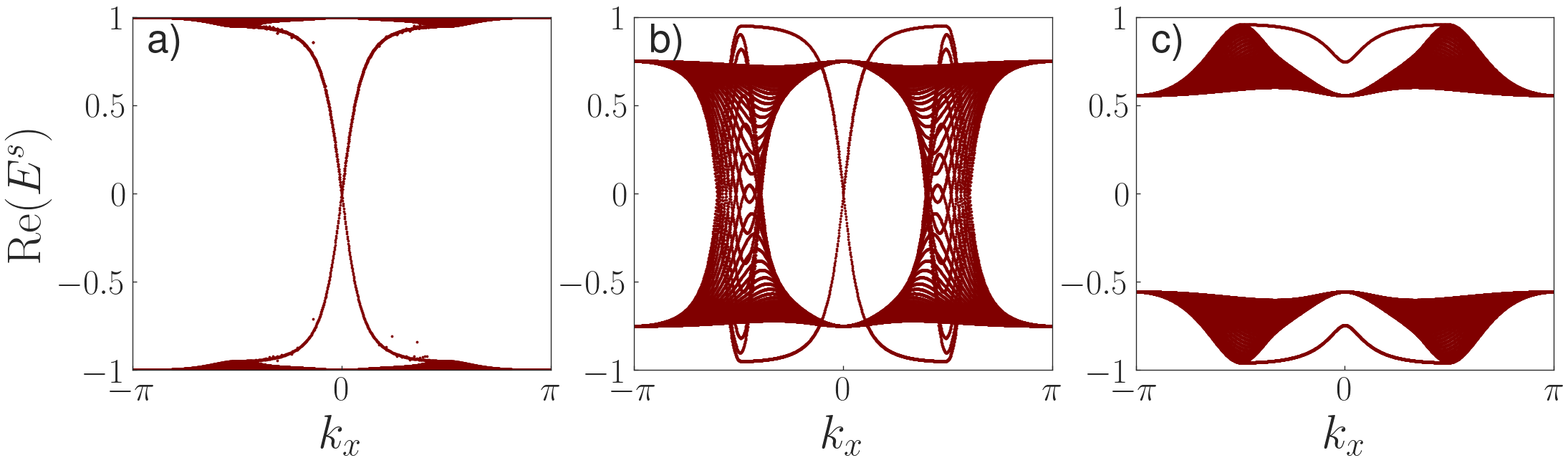}
\caption{The real energy spectra for the NH projected spin operator on the nanoribbon configuration are plotted (Eq. \eqref{E7}) corresponding to (a) $\gamma=0.0$, (b) $\gamma=0.1$ and (c) $\gamma=0.3$. The bands correspond to the energy eigenvalues $\mathrm{Re}(E^s_1)$ and $\mathrm{Re}(E^s_4)$, similar to the Hermitian scenario as mentioned in Eq. \eqref{E6}. The value of $B_y$ is kept fixed at $0.2$.}
\label{Fig4}
\end{figure}
Here, $|u_n^L(k_x)\rangle$ (eigenvector of $H_R^\dagger$) and $|u_n^R(k_x)\rangle$ (eigenvector of $H_R$) refer to the left and right occupied eigenstates for the nanoribbon configuration.
The orthonormality condition in this case is given by, $$\left<u_m^L|u_n^R\right>=\delta_{mn},$$ where $m$ and $n$ correspond to the band index \cite{PhysRevLett.121.026808}.

Let us now discuss Fig. \ref{Fig4} in detail.
We observe that when $\gamma=0$ and $B_y\neq 0$, the band gap of the projected spin spectra is absent, as illustrated in Fig. \ref{Fig4}(a), and it remains so as long as $\gamma\leq |B_y|$, depicted in Fig. \ref{Fig4}(b).
This observation, in conjunction with the non-trivial spin Berry curvature (Fig. \ref{Fig3}), indicates that the inclusion of the IPMF only establishes an $\uparrow$$\leftrightarrow$$\downarrow$ spin scattering at the edges of the nanoribbon, without harming the topology of the bulk.
This, in turn, causes the spin-filtered states at the edges of the BHZ model, in the presence of IPMF, to be gapped, as shown in Fig. \ref{Fig1}(a).
Note that the gap in the spin spectra is non-existent both in bulk and at the edges in Fig. \ref{Fig4}(b).
However, as the value of $\gamma$ exceeds $|B_y|$ in Fig. \ref{Fig4}(c), the spectrum becomes fully gapped.
We thus establish that the $\uparrow\;\leftrightarrow\;\downarrow$ scattering caused by the IPMF is suppressed as the non-hermiticity becomes the dominating factor in the system.
This opens the gap in the spin spectra, as shown in Fig. \ref{Fig4}(c), and retrieves the spin-filtered edge states (Fig. \ref{Fig1}(b)).

\begin{figure}
\includegraphics[width=\columnwidth]{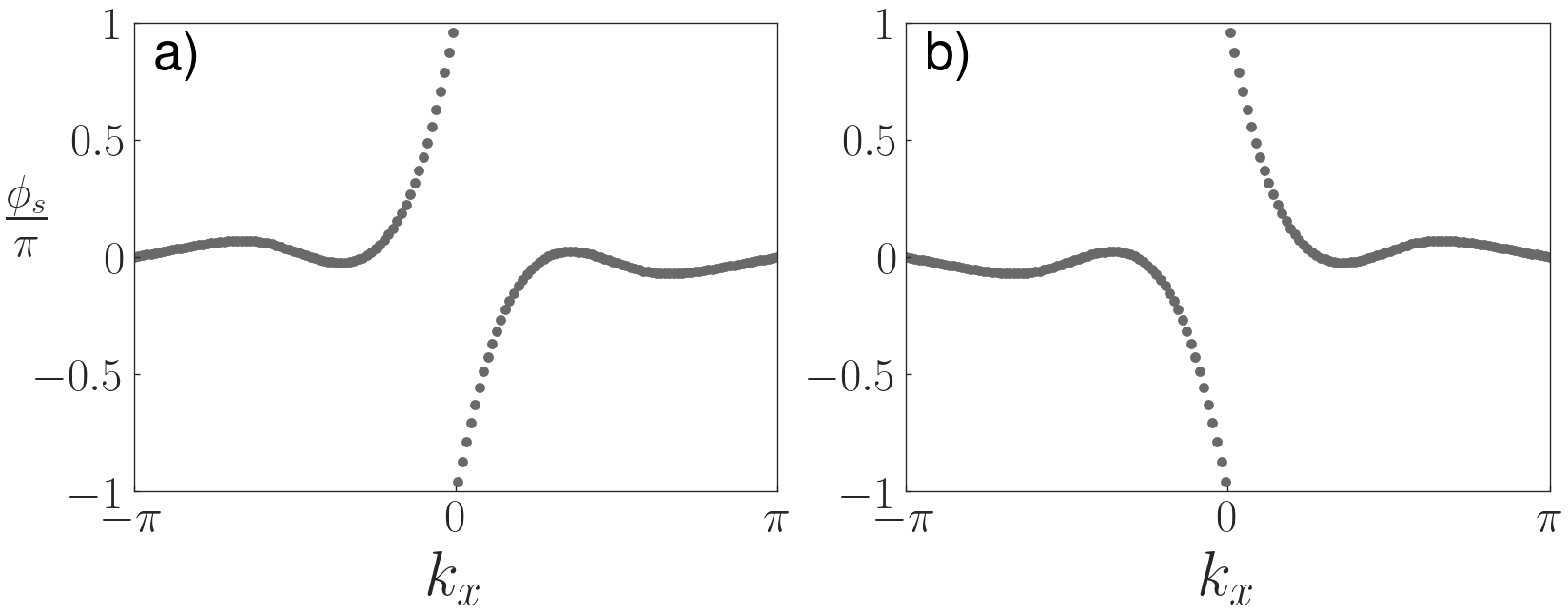}
\caption{The biorthoginal spin resolved Berry phase, corresponding to the projected (a) $\uparrow$-spin and (b) $\downarrow$-spin states, calculated using the eigenvectors $|\zeta_1^{L, R}\rangle$ and $|\zeta_4^{L, R}\rangle$ of the NH projected spin operator $S_\text{NH}(\mathbf{k}$), is plotted. $B_y$ and $\gamma$ bear the values $0.2$ and $0.3$, respectively.}
\label{Fig5}
\end{figure}

To further establish the topological nature of the edge states beyond $\gamma=|B_y|$, we plot the spin-resolved biorthogonal Berry phase calculated on a 1D closed loop along the $k_y$ direction as a function of $k_x$, illustrated in Figs. \ref{Fig5}(a) and \ref{Fig5}(b).
Physically, this quantity corresponds to the evolution of the charge center, as a function of $k_x$, for a spin-resolved hybrid Wannier function localized along the $y$-direction and periodic along $x$,
\begin{align}
\phi_s(k_x)=-\text{Im }\ln&\left[\text{G}_s(\mathbf{k_f}-\Delta{\mathbf{k}})\text{G}_s(\mathbf{k_f}-2\Delta{\mathbf k})...\right.\nonumber\\&\left.\text{G}_s(\mathbf{k_i}+\Delta{\mathbf k})\text{G}_s(\mathbf {k_i})\right]
\end{align}
where,
\begin{align}
\begin{split}
\text{G}_s({\mathbf{k}})&=\frac{1}{2}\Big\{\langle \zeta^{L}(\mathbf{k} + \Delta \mathbf{k})|\zeta^R(\mathbf{k})\rangle \\&+ \langle \zeta^R(\mathbf{k} + \Delta \mathbf k)|\zeta^L(\mathbf k)\rangle \Big \}
\end{split}
\end{align}
Here, $\Delta\mathbf k=\Delta k_y\hat y$ where $\Delta k_y=\frac{\mathbf{k_f}-\mathbf{k_i}}{N_y}$, corresponds to a small fraction of the one-dimensional closed loop ($\mathbf{k_i}\rightarrow\mathbf{k_f}$) along the $y$-direction divided into $N_y$ points \cite{PhysRevB.103.205110}.
$|\zeta^L\rangle$ and $|\zeta^R\rangle$ are the left and right eigenvectors obtained by diagonalizing the NH projected spin operator $S_\text{NH}(\mathbf{k})$ for a nanotube-configuration.
Again $\zeta^{L/R}\in\{ \zeta^{L/R}_1, \zeta^{L/R}_4\}$ similar to the Hermitian case as mentioned in Eq. \eqref{E6}, which in turn corresponds to the projected $\uparrow$-spin and $\downarrow$-spin states respectively.
In Fig. \ref{Fig5}, we observe a precise non-trivial winding of the spin-resolved Wannier charge center as a function of $k_x$ for both the projected $\uparrow$-spin (Fig. \ref{Fig5}(a)) and $\downarrow$-spin states (Fig. \ref{Fig5}(b)). 
This corresponds to the number of unit cells traversed by the center of charge in a complete cycle of the parameter $k_x$, thus equivalently indicating a non-trivial SCN in this regime \cite{PhysRevB.95.075146}.
Indeed, the non-Hermitian SCN is found to be $C_\uparrow=-C_\downarrow=1$ in the regime $\gamma>|B_y|$.
It is important to mention that any topological characterization based on the projected spin operator is not viable in the region where $\gamma\in(0, |B_y|]$.
This is because the bulk spectra of the spin operator $S_\text{NH}(\mathbf{k})$ remain gapless in this region, thus forbidding any meaningful discussion on the topological properties.

\section{Conclusion}

We establish that there exists a competition between two different orders of topological phase, namely the first and the second-order topological phase, in a non-Hermitian BHZ model in the presence of an in-plane magnetic field, controllable solely by the strength of a spin-dependent gain/loss potential.
The helical edge states for a $x$-periodic nanoribbon of the BHZ model, which were gapped out by the IPMF (thus giving rise to SOTP), are retrieved when the strength of the non-Hermiticity gains dominance over the IPMF.
We analyze this transition with the help of the spin Chern number calculated for the Hermitian case, which establishes that even in the presence of the IPMF, the first-order bulk topology remains unharmed.
On entering the NH regime, the spectrum of the NH projected spin operator for the nanoribbon configuration further shows that the IPMF, in the SOTP phase ($\gamma\leq|B_y|$), establishes channels that facilitate scattering between the $\uparrow$-spin and the $\downarrow$-spin states. 
This is the main reason for the appearance of a gap in the real energy spectrum of the nanoribbon-like configuration discussed by us.
The suppression of this spin channel by non-hermiticity is also exhibited by the spin spectrum of the nanoribbon in the regime $\gamma>|B_y|$, and subsequently, the corner modes in the rhombic supercell get delocalized along the edges.
This nuanced balance underscores the potential of non-Hermitian physics to manipulate and sustain topological phases of matter in environments where they would otherwise be destabilized by external perturbations.

\appendix
\section{\label{app1}$x$-periodic BHZ nanoribbon}
In this section, we are going to discuss a nanoribbon configuration of the Bernevig-Hughes-Zhang (BHZ) model, which is periodic in the $x$-direction, but has finite unit cells along the $y$-direction.
Before delving into the specifics of the nanoribbon, it is essential to elucidate the pillar symmetry inherent in the system, which corresponds to TRS.
Bernevig, Hughes, and Zhang considered a four-band model on a square lattice in which each unit cell contains two $s_{1/2}$ states, $\ket{s,\uparrow}$ and $\ket{s,\downarrow}$, and two $p_{3/2}$ states, $\ket{p_x+ip_y,\uparrow}$ and $\ket{p_x-ip_y,\downarrow}$.
In simplifying the notation of $p_{3/2}$ states corresponding to the $p$ orbital, we denote them as $\ket{p,\uparrow}$ for states with spin up, and $\ket{p,\downarrow}$ for states with spin down. 
The Hamiltonian of the BHZ model in the momentum space takes the form,
\begin{widetext}
    \begin{equation}
H_0(\pmb{k})=\sum_k\left(c^{\dagger}_{\pmb{k}s\uparrow}\;c^{\dagger}_{\pmb{k}p\uparrow}\;c^{\dagger}_{\pmb{k}s\downarrow}\;c^{\dagger}_{\pmb{k}p\downarrow}\right)H(\pmb{k})\begin{pmatrix}c_{\pmb{k}s\uparrow}\\c_{\pmb{k}p\uparrow}\\c_{\pmb{k}s\downarrow}\\c_{\pmb{k}p\downarrow}\end{pmatrix};\qquad H(\pmb{k})=\begin{pmatrix}
        h(\pmb{k}) & 0_{2\times 2}\\
        0_{2\times 2} & h^*(-\pmb{k})\end{pmatrix},
    \label{eq:Ham1}
\end{equation}
where 
\begin{equation}
    h(\pmb{k})=\begin{pmatrix}
    \Delta-2t(\cos{k_x}+\cos{k_y}) & 2t_{sp}(\sin{k_x}-i\sin{k_y})\\
    2t_{sp}(\sin{k_x}+i\sin{k_y}) & -\Delta+2t(\cos{k_x}+\cos{k_y})
\end{pmatrix}.
\end{equation}
\end{widetext}
$c_{\pmb{k}s\uparrow}$ ($c^{\dagger}_{\pmb{k}s\uparrow}$) denotes the annihilation (creation) operator for a spin-up ($\uparrow$) electron with a momentum $\pmb{k}$ in the $s$ orbital.
A similar notation has been adopted for the creation and annihilation operators corresponding to the $p$ orbitals.
$t$ and $t_{sp}$ denote (real-valued) hopping parameters for transitions between $s$ (or $p$) orbitals and $s$ to $p$ orbitals (or vice-versa) respectively, located in adjacent sites. 
$\Delta$ ($-\Delta$) represents the onsite potential for the $s$ ($p$) orbital.
The TRS operator is given by $\mathcal{T}=i\sigma_y\mathcal{K}\otimes\tau_0$, where $\sigma_y$ and $\tau_0$ act on the spin and the orbital bases, respectively.
$\mathcal{K}$ is simply the complex conjugation operator.
The TRS in the BHZ model is verified through the relation, $\mathcal{T}H^*(-\pmb{k})\mathcal{T}^{-1}=H(\pmb{k})$, where $H(\pmb{k})$ is the Bloch Hamiltonian in Eq. \eqref{eq:Ham1}.
When an in-plane magnetic field (IPMF), $\pmb{B}$, oriented in the  $x$-$y$ plane is applied to the system, TRS is inherently broken due to the Zeeman term.
This breaking of TRS typically destroys the quantum spin Hall (QSH) states, which crucially rely on the TRS for their existence and stability.
However, including a spin-dependent imaginary potential, represented by $H_\text{NH} = \mathrm{i}\gamma\sigma_z\otimes\tau_0$, to $H(\pmb{k})$ adds a layer of complexity.
Unlike real potentials that can perturb the energy levels of a system, staggered imaginary potentials that are dependent upon the spin state of the electrons introduce non-Hermitian (NH) effects, rendering gain/loss properties to the system.
The key aspect of incorporating non-Hermiticity is that it can be engineered to preserve the TRS of the composite system.
Or in other words, the Hamiltonian, $H(\pmb{k})+H_{\text{NH}}$, is TR symmetric.

Now, we shall discuss the $x$-periodic BHZ nanoribbon.
The total number of the unit cells in the $y$-direction is $L_y$, and the Hamiltonian of the nanoribbon is given by
\begin{widetext}
    \begin{align*}
        H(n_y, k_x)=\sum_{n_y,k_x,\sigma,\alpha}&(-1)^\alpha\Delta c^{\dagger}_{n_y,k_x\alpha\sigma}c_{n_y,k_x\alpha\sigma}-\sum_{n_y,k_x,\sigma}\left[2t\cos{k_x}\left(c^{\dagger}_{n_y,k_xs\sigma}c_{n_y,k_xs\sigma}-c^{\dagger}_{n_y,k_xp\sigma}c_{n_y,k_xp\sigma}\right)\right.\nonumber\\&\left.+t\left(c^{\dagger}_{n_y+1,k_xs\sigma}c_{n_y,k_xs\sigma}-c^{\dagger}_{n_y+1,k_xp\sigma}c_{n_y,k_xp\sigma}\right)+t_{sp}\left(c^{\dagger}_{n_y+1,k_xs\sigma}c_{n_y,k_xp\sigma}-c^{\dagger}_{n_y-1,k_xs\sigma}c_{n_y,k_xp\sigma}\right)+\right.\nonumber\\&\left.\hspace{8cm}2it_{sp}\sin{k_x}e^{\frac{i\sigma\pi}{2}}c^{\dagger}_{n_y,k_xs\sigma}c_{n_y,k_xp\sigma}\right]+\text{H.c.},
\end{align*}
\end{widetext}
where $\sigma$ denotes the spin orientation, taking a numerical value of $+1$ ($-1$) corresponding to the up (down) spin and $\alpha$ represents the orbital type within a unit cell, that is, $\alpha\equiv(s,p)$, and takes the values, $0, 1$, corresponding to the $s$, $p$ orbitals respectively.
Furthermore, $c_{n_y,k_x\alpha\sigma}$ ($c^{\dagger}_{n_y,k_x\alpha\sigma}$) is the annihilation (creation) operator for an electron with spin, $\sigma$ (up or down), in orbital $\alpha$ of the $n_y^{\text{th}}$ unit cell in the $y$-direction, possessing a momentum $k_x$.
$n_y$ runs from $1$ to $L_y$.
When $\Delta/2t < 2$, a pair of edge states appear along the boundary of the system, as shown in Fig. \ref{fig:1s}(a).
These edge states remain robust even when subjected to TR invariant perturbations and are protected by the presence of TRS, which gives rise to a $Z_2$ topological invariant.
Consequently, this system is equivalent to two independent quantum Hall systems with equal and opposite Hall conductivities, which guarantees a pair of spin-filtered counter-propagating edge states on each boundary of the system.
In essence, the system represents a topologically nontrivial insulator that cannot be smoothly transformed into a trivial insulator through adiabatic processes.
\begin{widetext}
\onecolumngrid
\begin{figure}[t]
    \centering
    \includegraphics[width=\columnwidth, height=0.28\columnwidth]{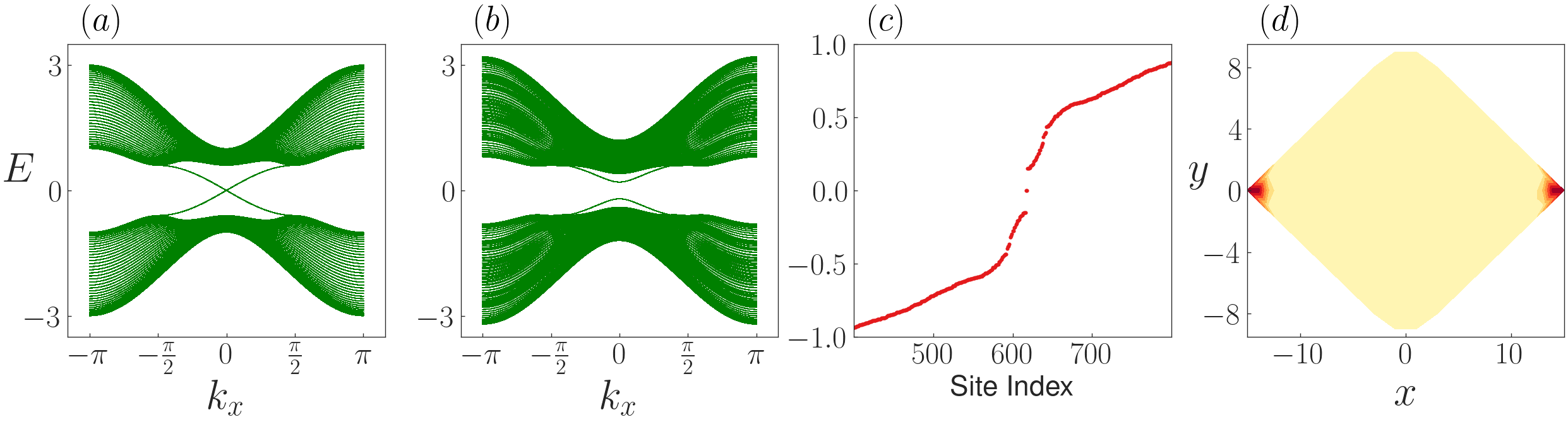}
    \caption{The spectra of the Hermitian $x$-periodic nanoribbon, $H(n_y, k_x)$, is depicted in (a) the absence ($|B_y|=0$), and (b) the presence ($|B_y|=0.2$) of the IPMF. Without the IPMF, the system resembles the original BHZ model with helical edge states. However, on applying the IPMF, these helical edge states gap out. (c) The energy spectra of the Hermitian BHZ model under the influence of the IPMF ($|B_y|=0.2$) show the presence of distinct zero energy states separated from the bulk when projected on a rhombic supercell. (d) The probability density of the zero energy states shows confinement at two specific corners of the rhombic supercell $|B_y|=0.2$.}
    \label{fig:1s}
\end{figure}
\begin{figure}[h]
    \centering
    \includegraphics[width=0.7\columnwidth, height=0.35\columnwidth]{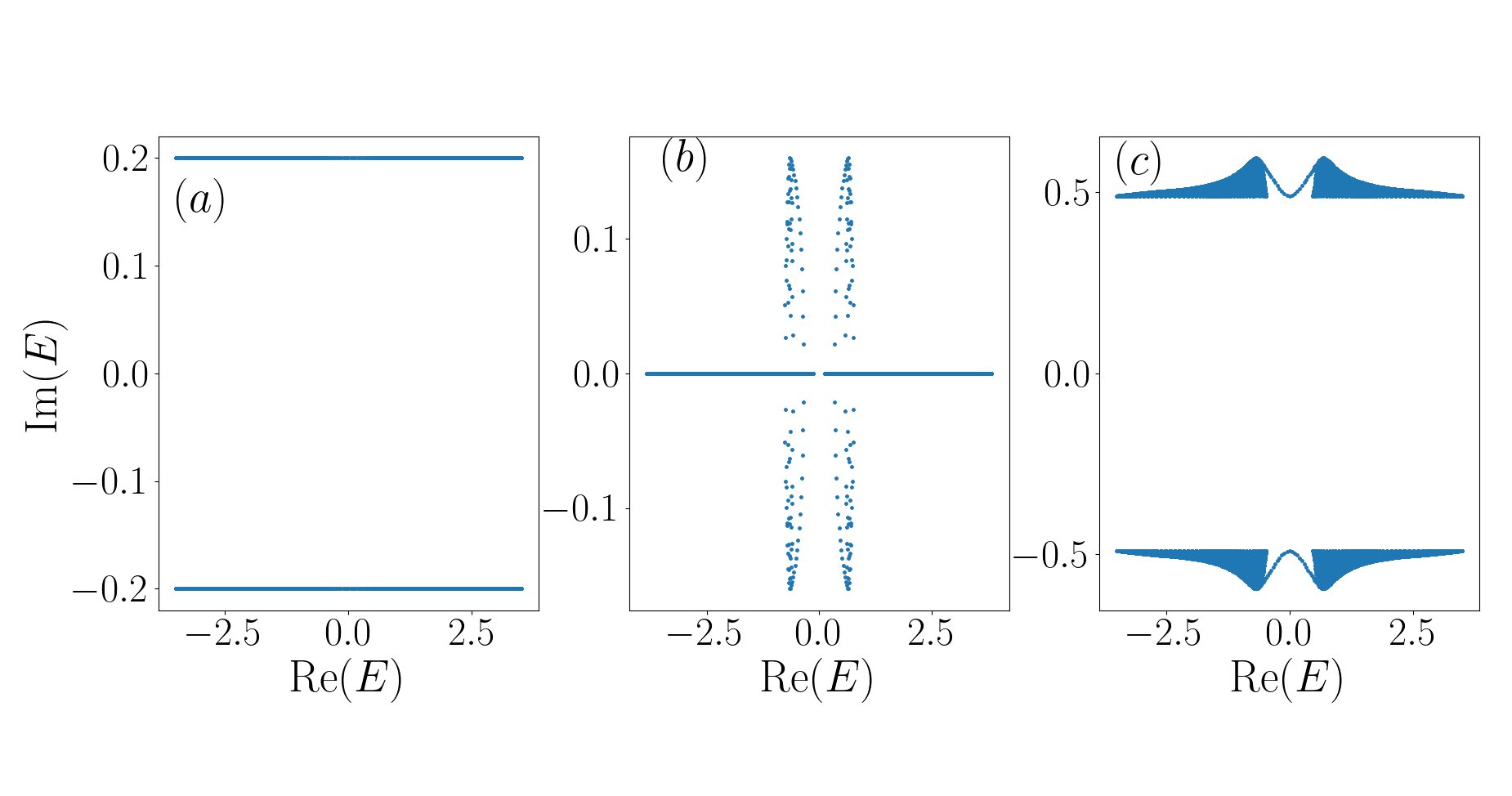}
    \caption{Real vs imaginary parts of the eigenspectra of an $x$-periodic ribbon are plotted for (a) $|B_y|=0,\gamma=0.2$, (b) $|B_y|=0.4,\gamma=0.2$, and (c) $|B_y|=0.5,\gamma=0.7$.}
    \label{fig:2s}
\end{figure}
\end{widetext}

The introduction of the IPMF, described by $H_\text{Z}=B_y\sigma_y\otimes\tau_0$, to the Hamiltonian, $H(n_y,k_x)$, fundamentally alters the symmetry properties of the system.
The direct consequence of breaking TRS with an IPMF is the destabilization and eventual vanishing of the QSH states (Fig. \ref{fig:1s}(b)) and the emergence of two pairs of zero energy in gap corner states, which is vividly captured in the associated energy spectra for a rhombic supercell with fully OBC, shown in Fig. \ref{fig:1s}(c).
These states are found to be localized at the two opposite corners of the rhombic supercell, as shown in Fig. \ref{fig:1s}(d), and provide a clear illustration of the second-order topological phase (SOTP).
Unlike the helical edge states, SOTPs are characterized by states that are localized at corners of 2D supercells (or hinges in 3D) of the system.
This transformation from edge states to corner states under the influence of an IPMF is a clear demonstration of how external fields can be utilized to manipulate the transition between different topological phases.

Now, let us include the spin-dependent imaginary potential, $H_{\text{NH}}$, to the Hamiltonian $H(n_y, k_x)$ and study the eigenspectra for the nanoribbon in the absence of the IPMF.
As established earlier, the inclusion of $H_{\text{NH}}$ does not disrupt the TRS of the system, thus ensuring the stability of QSH states.
However, $H_{\text{NH}}$ introduces an imaginary component to the energy spectrum, leading to all eigenenergies appearing as pairs of $E$ and $E^*$ ($=E\pm i\gamma$), as illustrated in Fig. \ref{fig:2s}(a), as a manifestation of the TRS in the system.
The absence of an energy gap along the real axis, coupled with a gap of $2\gamma$ along the imaginary axis, signifies that the erstwhile robust helical edge states have now become dynamically unstable.
The introduction of the IPMF into the NH system brings about a nuanced interplay between the TRS-breaking effects due to the IPMF and the TRS-preserving influence of $H_\text{NH}$.
The underlying mechanism hinges on the ability of $H_\text{NH}$ to modify the energy landscape of the system in such a way that it mimics the conditions necessary for QSH states to emerge, effectively mitigating the disruptive influence of the IPMF.
As depicted in Fig. \ref{fig:2s}(b), when $|B_y|>\gamma$, an energy gap emerges along the real axis, indicating the absence of zero-energy helical edge states.
Conversely, in the regime where $|B_y|<\gamma$, as illustrated in Fig. \ref{fig:2s}(c), the dominance of $H_{\text{NH}}$ over the IPMF leads to the reassurance of QSH states.
In this case, an energy gap appears along the imaginary axis, suggesting a transition from a real line gap to an imaginary line gap transition at $|B_y|=\gamma$.
This transition underscores the critical role played by the interplay between the competing influences of $H_\text{NH}$ and $B_y$ in shaping the topological characteristics of the system.

\section{\label{app2}$y$-periodic BHZ nanoribbon}
Let us discuss the $y$-periodic BHZ nanoribbon, with a total number of unit cells in the $x$-direction being $L_x$.
The Hamiltonian of the nanoribbon is given by,
\begin{widetext}
\onecolumngrid
    \begin{align*}
    H(n_x,k_y)=&\sum_{n_x,k_y,\sigma,\alpha}(-1)^\alpha\Delta c^{\dagger}_{n_x,k_y\alpha\sigma}c_{n_x,k_y\alpha\sigma}-\sum_{n_x,k_y,\sigma}\left[2t\cos{k_y}\left(c^{\dagger}_{n_x,k_ys\sigma}c_{n_x,k_ys\sigma}-c^{\dagger}_{n_x,k_yp\sigma}c_{n_x,k_yp\sigma}\right)+\right.\nonumber\\&\left.t\left(c^{\dagger}_{n_x+1,k_ys\sigma}c_{n_x,k_ys\sigma}-c^{\dagger}_{n_x+1,k_yp\sigma}c_{n_x,k_yp\sigma}\right)+t_{sp}e^{\frac{i\sigma\pi}{2}}\left(c^{\dagger}_{n_x+1,k_ys\sigma}c_{n_x,k_yp\sigma}-c^{\dagger}_{n_x-1,k_ys\sigma}c_{n_x,k_yp\sigma}\right)+\right.\nonumber\\&\left.\hspace{9cm}2it_{sp}\sin{k_y}c^{\dagger}_{n_x,k_ys\sigma}c_{n_x,k_yp\sigma}\right]+\text{H.c.},
\end{align*}
\begin{figure}[h]
    \centering
    \includegraphics[width=0.9\columnwidth, height=0.27\columnwidth]{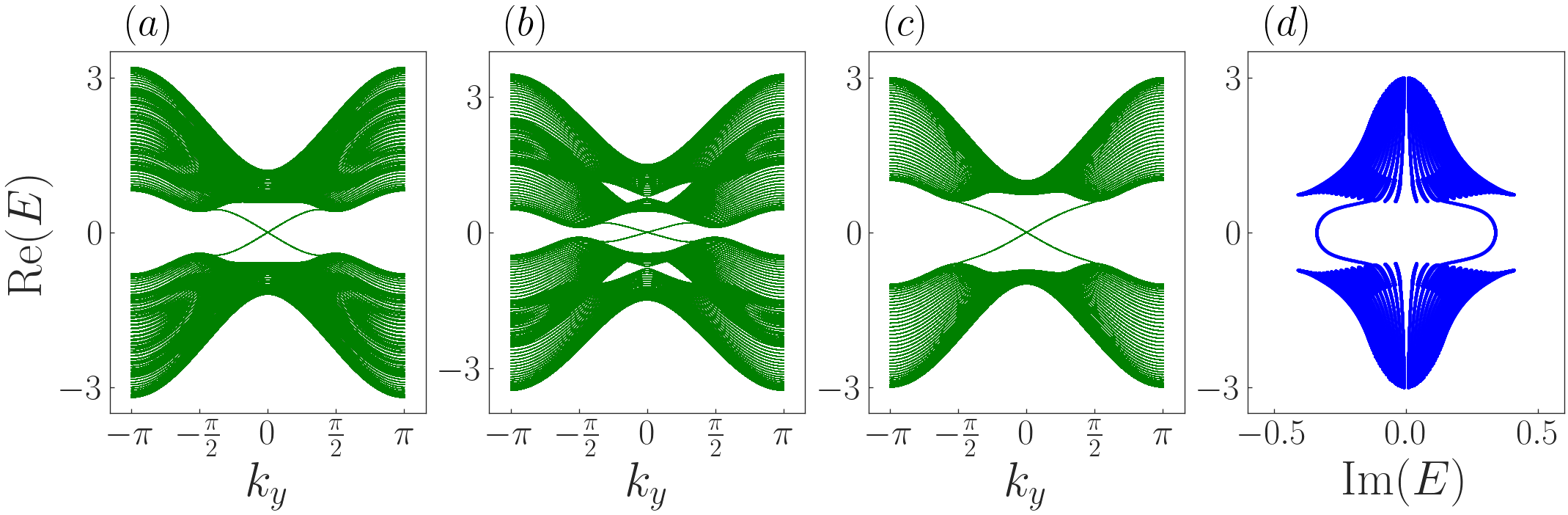}
    \caption{Band structure of $H(n_x,k_y)$ in the presence of IPMF is plotted for (a) $|B_y|=0.2$, (b) $|B_y|=0.5$ in the absence of any non-Hermiticity. It is observed that the introduction of the IPMF interestingly generates duplicates of each energy band and narrows the gap between these bands, steering the system closer to a (semi-)metallic state. Introducing a spin-dependent imaginary potential prevents the overlap of bands, thereby preserving the existence of the QSH states as seen in (c) where $|B_y|=\gamma=0.5$. Subsequently, in panel (d), the real and imaginary components of the energy spectra depicted in panel (c) are plotted. These spectra reveal that the QSH states observed in panel (c) are dynamically unstable, as evidenced by their complex energy values. The total number of unit cells in the $x$-direction is fixed at $32$. The other parameters are $t=0.5$, $t_{sp}=0.3$, and $\Delta=1.0$.}
    \label{fig:3s}
\end{figure}
\end{widetext}
where the mathematical symbols bear usual meaning as those for the $x$-periodic BHZ nanoribbon.
Let us now discuss the fate of QSH states in the presence of the IPMF.
Despite the disruption of time-reversal symmetry by $B_y$, the unique helical edge states show resilience at lower magnetic field strengths, as demonstrated in Fig. \ref{fig:3s}(a).
The magnetic field does influence the system by causing a spin split and narrowing the band gap.
As the strength of $B_y$ increases, we observe the bands drawing closer, leading to an eventual overlap shown in Fig. \ref{fig:3s}(b).
This signals a shift towards a more (semi-)metallic state with the formation of sub-bands.
However it can be safely said that the effect of IPMF on the edge states of a $y$-periodic nanoribbon is much weaker as compared to an $x$-periodic one.

When $H_\text{NH}$ is incorporated into the Hamiltonian $H(n_x,k_y)$ in the absence of the IPMF, the Hamiltonian exhibits results analogous to those observed in the $x$-periodic NH BHZ nanoribbon, as particularly noted in Fig. \ref{fig:2s}(a).
Upon integrating the IPMF with the NH BHZ nanoribbon, intriguing characteristics unfold within the system.
As illustrated in Fig. \ref{fig:3s}(c), non-Hermiticity effectively resists the impending band overlap initially caused by the IPMF (as depicted in Fig. \ref{fig:3s}(b)), leading to the re-strengthening of the QSH states.
This scenario sets the stage for a rivalry between the NH potential, $H_\text{NH}$, and the IPMF, $B_y$.
$H_\text{NH}$ strives to uphold the topological integrity of the system fostering the QSH phase and its associated edge states, while $B_y$ induces a (semi-)metallic behavior in the system.
Understanding this intricate balance between these competing factors is crucial for deciphering the system's behavior and predicting its electronic properties.
The real and imaginary components of the eigenvalues for the Hamiltonian are demonstrated in Fig. \ref{fig:3s}(d).
Similar to the scenario observed in the $x$-periodic BHZ nanoribbon, the inclusion of $H_\text{NH}$ introduces dynamical instability to the QSH states.
This highlights a nuanced interplay between the non-Hermiticity and external magnetic fields, revealing that, while non-Hermiticity can counteract the effects of TRS-breaking fields to preserve topological states, it simultaneously introduces an element of instability to these states.

\bibliographystyle{unsrtnat}
\bibliography{ref3.bib}

\end{document}